\documentclass{appolb}

\usepackage{amsmath}
\usepackage[figuresright]{rotating}
\usepackage{lineno}

\begin{document}
%\linenumbers

\title{Higher-order Proton Cumulants in Au+Au Collisions at $\sqrt{s_{\rm NN}}$ = 3 GeV from RHIC-STAR\thanks{Presented at Quark Matter 2022, Karkow, Poland, April 4-10, 2022}}

\author{Yu Zhang \\for the STAR Collaboration
\address{yuz@ccnu.edu.cn}
\address{Key Laboratory of Quark \& Lepton Physics (MOE) and Institute of Particle Physics, Central China Normal University, Wuhan, 430079, China}
}
\maketitle
\begin{abstract}
%The higher-order fluctuations of conserved quantities such as net baryon number are predicted to be sensitive to the non-equilibrium correlation length, $\xi$, and thus serve as indicators of critical behavior. Experimentally, fluctuations of proton and anti-proton numbers have been shown to be reliable proxies for baryons and anti-baryons. In the first Beam Energy Scan (BES-I) at the Relativistic Heavy Ion Collider (RHIC), which was run from 2010 to 2014, the higher-order cumulant ratio, $C_4/C_2$, of the net-proton multiplicity distributions shows a non-monotonic energy dependence between the energies of 7.7 to 200 GeV with a significance of 3.1$\sigma$. Motivated by the findings of BES-I, the Solenoidal Tracker at RHIC (STAR) collaboration improved the detector performance of the STAR detector and began two additional physics programs: the BES-II and the fixed-target (FXT) program. While BES-II revisits the energies of BES-I with higher statistics and improved detector performance, the FXT program extends the lowest energy from $\sqrt{s_{\rm NN}}$ = 7.7 GeV to $\sqrt{s_{\rm NN}}$ = 3 GeV.

In these proceedings, we present the higher-order cumulants of proton multiplicity distributions of the fixed-target (FXT) run in Au+Au collisions at $\sqrt{s_{\rm NN}}$ = 3.0 GeV. The cumulant ratios are presented as a function of centrality and collision energy. The proton cumulant ratio $C_4/C_2$ is consistent with fluctuations driven by baryon number conservation and indicates an energy regime dominated by hadronic interactions. These data imply that the QCD critical point could exist at energies higher than 3 GeV if created in heavy-ion collisions. 
%We discuss the physics implications of these results with comparisons to results from the HADES experiment and a hadronic transport model.
\end{abstract}

%% main text
\section{Introduction}
Experimental evidences~\cite{
%BRAHMS:2004adc,PHOBOS:2004zne,PHENIX:2004vcz,
STAR:2005gfr} at RHIC and the LHC have demonstrated the formation of Quark-Gluon Plasma (QGP) in ultra-relativistic heavy-ion collisions at small baryon chemical potential ($\mu_{B} \approx 0$ MeV) where the phase transition from the hadronic matter to QGP is suggested to be a crossover from state-of-the-art Lattice QCD calculations~\cite{Aoki:2006we}. It has been conjectured that there is a first-order phase transition and a QCD critical point at the finite $\mu_{B}$ region in the QCD phase diagram. In the search for the possible QCD critical point, higher-order cumulants of conserved quantities such as net-baryon number, net-strangeness number, and net-charge number are sensitive observables to locate its position~\cite{Stephanov:2008qz,Asakawa:2009aj,Luo:2017faz,Gupta:2011wh}. Experimentally net-proton and net-kaon numbers are used as a proxy for net-baryon and net-strangeness numbers due to the difficulty to detect neutral particles in the experiment. Recent results from the STAR experiment on net-proton fourth-order cumulant ratio have shown intriguing non-monotonic energy dependence with 3.1\,$\sigma$ significance in the most central Au+Au collisions at $\sqrt{s_{\rm NN}}$ =  7.7 - 200 GeV~\cite{STAR:2020tga,STAR:2021iop} while there are still large statistical uncertainties for energy $\sqrt{s_{\rm NN}}<19.6$ GeV.
% and an energy gap below $\sqrt{s_{\rm NN}}$ = 7.7 GeV. 
 These proceedings reports proton cumulants and cumulant ratios up to $4^{\rm th}$-order in $\sqrt{s_{\rm NN}}$ = 3 GeV Au+Au collisions from the STAR fixed-target experiment. The relevant analysis details and correction methods will also be shortly discussed. To understand the collision dynamics in the absence of the critical behavior, we have carried out simulations with a microscopic transport model UrQMD~\cite{Bleicher:1999xi} for Au+Au collisions at $\sqrt{s_{\rm NN}}$ = 3 GeV. Connections between experimental data and physics implications in the high baryon density region will be discussed.

\section{Experimental Observables}
This section shows the definitions of cumulants and cumulant ratios. Let $N$ represent net-proton number. The deviation from its mean value ($\langle N\rangle$) is defined as $\delta N=N-\langle N\rangle$. Then cumulants up to $4^{\rm th}$-order can be written as:
\begin{equation}
\begin{split}
C_{1} &= \langle N\rangle,
C_{2} = \langle (\delta N)^{2}\rangle,
C_{3} = \langle (\delta N)^{3}\rangle,\\
C_{4} &= \langle (\delta N)^{4}\rangle-3\langle (\delta N)^{2}\rangle^{2}.
\end{split}
\end{equation}
The cumulants are related to the various moments as 
\begin{equation}
	 \begin{split}
	M = C_1, \quad \sigma^2 = C_2, \quad S = \frac{C_3}{(C_2)^{3/2}}, \quad \kappa = \frac{C_4}{C_2^2},
	\end{split} 
\end{equation} where $M$, $\sigma^2$, $S$, and $\kappa$ are mean, variance, skewness, and kurtosis, respectively. Various cumulant ratios like $C_2/C_1$, $C_{3}/C_{2}$, and $C_{4}/C_{2}$ are constructed to cancel volume dependence: 
\begin{equation}
	\frac{C_2}{C_1} = \sigma^2/M, \quad\frac{C_3}{C_2}=  S\sigma, \quad \frac{C_4}{C_2}=\kappa\sigma^2.
\end{equation}

\section{Analysis Details}
\begin{figure}[htbp]
\begin{center}
\includegraphics[width=1.0\textwidth]{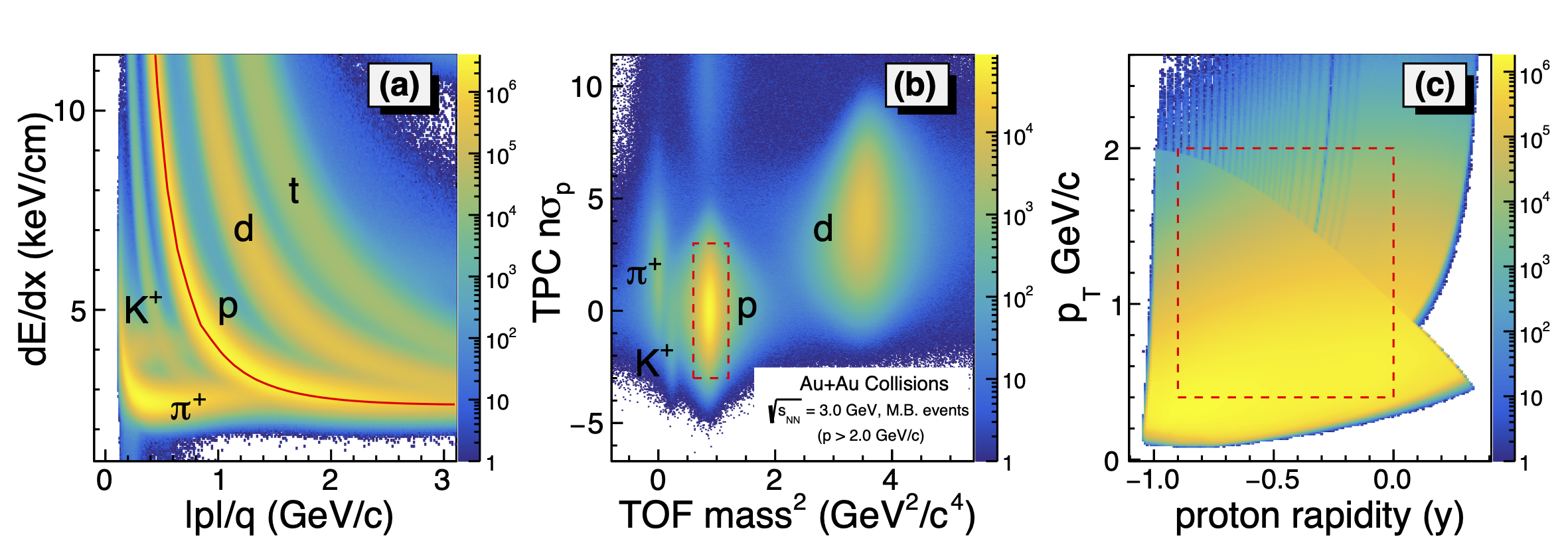}
\caption{Panel (a): TPC track energy loss (dE/dx (keV/cm)) vs. momentum; pion, kaon, deuteron and triton are labeled.  The proton Bethe-Bloch curve is plotted with red line. Panel (b): TPC $n\sigma_{p}$ vs. TOF $\rm mass^{2}$. Panel (c): Transverse momentum ($p_{\rm T}$) vs. proton rapidity.
}
\label{fig:pid_plot}
\end{center}
\end{figure}

The analysis used around 140 million $\sqrt{s_{\rm NN}}$ = 3 GeV Au+Au collisions events which are collected by the dedicated physics fixed-target run of the STAR experiment in the year 2018. The centrality is determined using charged particle reference multiplicity excluding protons and light nuclei within $-2 < \eta < 0$ where $\eta$ is pseudo-rapidity in the lab frame. As shown in Fig.~\ref{fig:pid_plot} protons are identified  by comparing the enegy loss measured by the Time Projection Chamber (TPC) with theoretical predictions (Fig.~\ref{fig:pid_plot}(a)). At high momentum ($p_{\rm lab}>2$ GeV/$c$), due to the contamination from other particles,  the mass square measured by Time of Flight (TOF) is used to ensure proton purity (Fig.~\ref{fig:pid_plot}(b)). The anti-protons are negligible ($\bar{p}/p < 10^{-6}$) at $\sqrt{s_{\rm NN}}$ = 3 GeV thus the proton cumulants are measured in the analysis. Figure~\ref{fig:pid_plot}(c) shows proton acceptance with the combination of TPC and TOF. The red dashed box indicates the acceptance window used in this analysis.

Cumulants are corrected for detector efficiencies by a track-by-track method~\cite{Nonaka:2017kko,Luo:2018ofd}. The rapidity (y) and transverse-momentum ($p_{\rm T}$) dependences of detector efficiency are considered. To correct the pileup effect due to the finite thickness of the gold target, a pileup correction method~\cite{Nonaka:2020qcm,Zhang:2021rmu} is used. As seen in our model simulation, there is a large initial volume fluctuation effect when calculating cumulants at $\sqrt{s_{\rm NN}}$ = 3 GeV, thus we tested an initial volume fluctuation correction method~\cite{Braun-Munzinger:2016yjz}. We measured cumulants as a function of reference multiplicity, and then obtained centrality binned results by the Centrality Bin Width Correction (CBWC)~\cite{Luo:2013bmi}. The statistical uncertainties of cumulants are estimated by the bootstrap method. The systematic uncertainties are estimated by varying analysis cuts related to centrality, pileup effect, track quality, and detector efficiency.%~\cite{bootstrap_method}.%But due to its strong model dependence, the correction was not used in the final result.
%In this year 2018, STAR collected around 250 million events with the fixed-target experiment in Au+Au collision at $\sqrt{s_{\rm NN}}=3$ GeV. 3 GeV is the lowest energy point from the STAR fixed-target experiment. The net-proton fluctuation measurements at this energy will enable us to discover the QCD phase diagram in a wide baryon chemical potential range.

%\begin{figure}[htbp]
%\begin{center}
%\includegraphics[width=0.7\textwidth]{refmult_v3.pdf}
%\caption{Panel~(a): TPC Track energy loss (dE/dx (KeV/$cm$)) vs. momentum; pion, kaon, deuteron and triton are labeled. The proton Bethe-Bloch curve is plotted with red line.  Panel~(b): TPC $n\sigma_{p}$ vs. TOF $mass^{2}$. Panel~(c): Transverse momentum ($p_{\mathrm{T}}$) vs. proton rapidity. }
%\end{center}
%\end{figure}
%\label{fig:pid}

\begin{figure}[htbp]
\begin{center}
\includegraphics[width=0.65\textwidth]{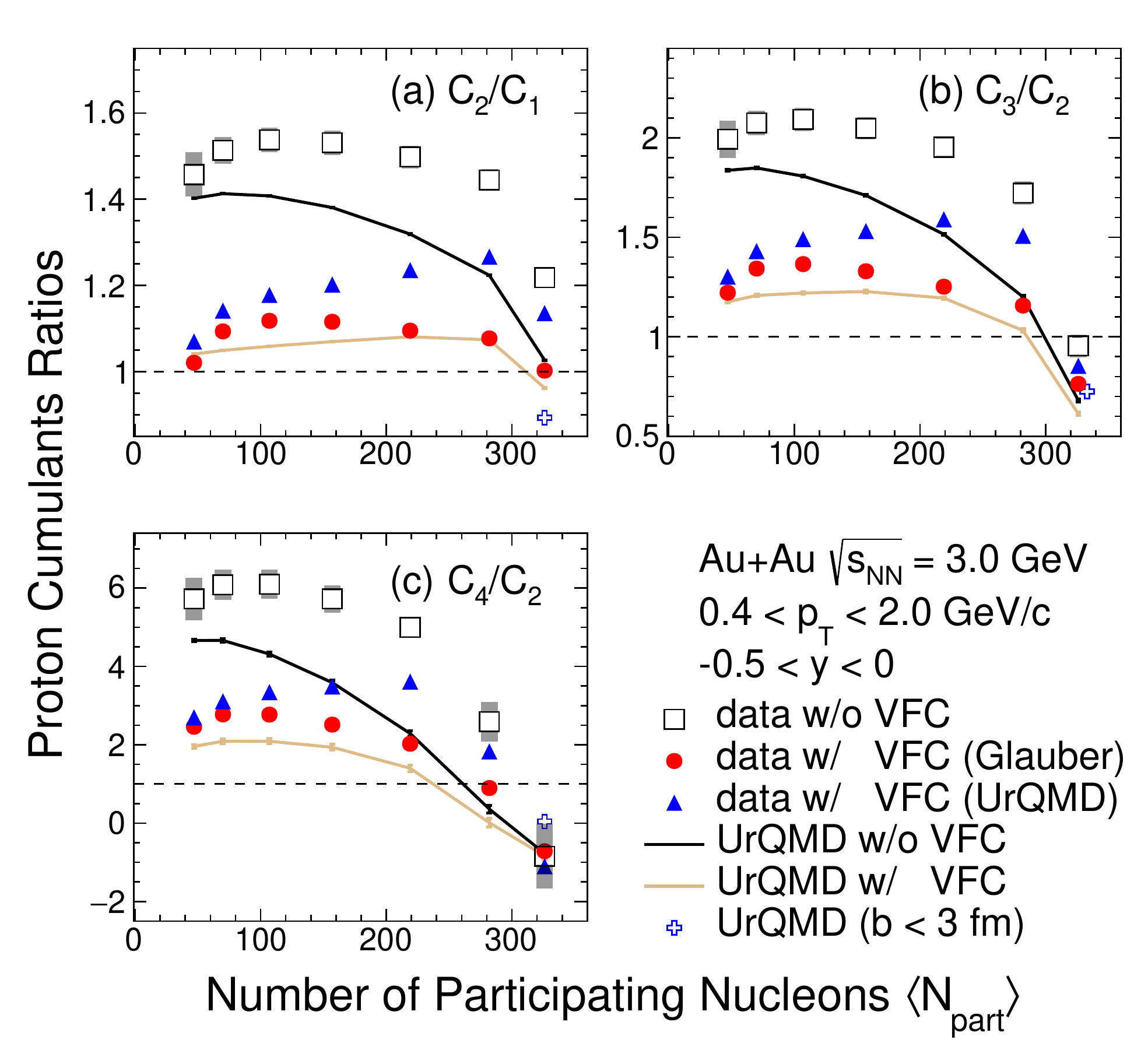}
\caption{Centrality dependence of proton cumulants and cumulant ratios up to $4^{\rm th}$-order in Au+Au collisions at $\sqrt{s_{\rm NN}}$ = 3\,GeV within kinematic acceptance $-0.5 < y < 0$ and $0.4 < p_{\mathrm{T}} < 2.0$ GeV/$c$. The black squares are results without volume correction while red circles and blue triangles represent results with volume correction using Glauber and UrQMD model, respectively.}
\label{fig:centrality}
\end{center}
\end{figure}

%
%
%\begin{figure}[htbp]
%\begin{center}
%\includegraphics[width=0.7\textwidth]{fig_upto4_v2.pdf}
%\caption{Centrality dependence of cumulants and cumulant ratios of proton multiplicity distributions up to $6^{\rm th}$-order in Au+Au collisions at $\sqrt{s_{\rm NN}}$ = 3\,GeV within kinematic acceptance $-0.5 < y < 0$ and $0.4 < p_{\mathrm{T}} < 2.0$ GeV/$c$. Data are shown with black squares while UrQMD results are shown with gold band. Statistical and systematical uncertainty are shown with black and grey bars, respectively.}
%\end{center}
%\end{figure}
%\label{fig:pid}

%\begin{figure}[htbp]
%\begin{center}
%\includegraphics[width=0.5\textwidth]{fig_Rapidity1.pdf}
%\caption{Rapidity dependence of cumulants and cumulant ratios of proton multiplicity distributions ratios up to $6^{\rm th}$ order in top 5\% central and 50-60\% peripheral Au+Au collisions at $\sqrt{s_{\rm NN}}$ = 3\,GeV within kinematic acceptance $0.4 < p_{\mathrm{T}} < 2.0$ GeV/$c$. The lower rapidity cut is varied from -0.1 to -0.9. Data are shown with black squares while UrQMD calculations are shown with gold bands. Statistical and systematical uncertainty are shown with black and grey bars, respectively.}
%\label{fig:rapidity}
%\end{center}
%\end{figure}

\begin{figure}[htbp]
\begin{center}
\includegraphics[width=0.65\textwidth]{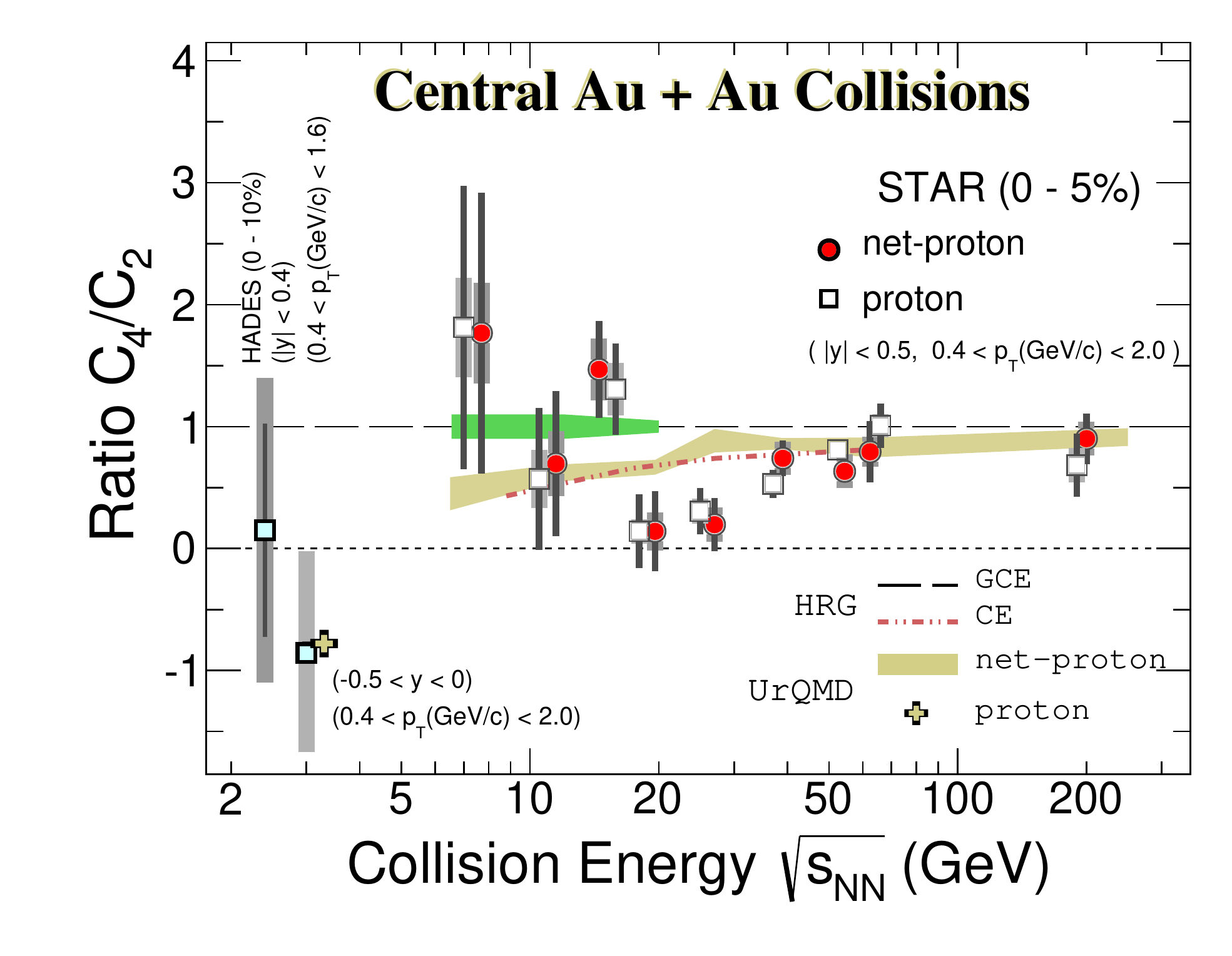}
\caption{Collision energy dependence of cumulants ratio $C_4/C_2$ in central Au+Au collisions within kinematic acceptance cut $0.4 < p_{\mathrm{T}} < 2.0$ GeV/$c$.
%Results from STAR data and UrQMD within $|y|<0.5$ are shown with red filled circles for net-proton, black squares for proton. 
%Results from data and UrQMD model with rapidity cut $ -0.5 < y < 0$ are shown with filled squares and gold filled crosses, respectively.
%UrQMD calculation with rapidity cut $|y|<0.5$ is shown with gold band for net-proton.
UrQMD calculations are shown with gold band for net-proton with rapidity cut $|y|<0.5$ and gold filled cross for proton with rapidity cut $-0.5<y<0$. 
%The result from HADES collaboration with rapidity cut $|y|<0.4$ in 0-10\% Au+Au collisions at $\sqrt{s_{\rm NN}}$ = 2.4 GeV is shown with cyan filled square. A calculation from the Hadron Resonance Gas (HRG) Model with grand canonical ensemble and canonical ensemble are shown with black dashed line and red dashed line, respectively. 
Statistical and systematic uncertainty are shown with black and grey bars, respectively. The green shaded area indicates the projected statistical uncertainty with BES-II data.}
\label{fig:energy}
\end{center}
\end{figure}

\section{Results}
Figure \ref{fig:centrality} shows the centrality dependence of the proton cumulant ratios $C_2/C_1$, $C_3/C_2$, and $C_4/C_2$ within $-0.5<y<0$ and $0.4<p_{\rm T}<2.0$ GeV/$c$. The 3 GeV data shown with black open squares are corrected for detector efficiency and pileup effect and then the CBWC was applied to obtain centrality binned results. The red circles and blue triangles are additionally corrected for initial volume fluctuation using Glauber and UrQMD models, respectively.
%From the comparison of the data points, the volume fluctuation correction shows strong model dependence while the most central centrality is least affected especially for higher-order ratio $C_4/C_2$. The gold and black lines are calculations from the UrQMD model with and without the volume fluctuation correction. 
%The comparison within UrQMD calculations supports the conclusion that $C_4/C_2$ in most central centrality bin is least affected by the volume fluctuation correction.
It is clear that the volume fluctuation correction shows a strong model dependence and affects the centrality dependence, particularly in peripheral collisions. The respective dynamics in the UrQMD and Glauber model for charged hadron production lead to two different mappings from the measured final charged hadron multiplicity distributions to the initial geometry. This difference is likely to be the dominant source of the model dependence in the VFC. On the other hand, one can see in the figure that the difference between results with and without the VFC is small for higher order ratios $C_3/C_2$ and $C_4/C_2$ in the most central bin.
% Centrality dependence of the proton cumulant ratios 
%The blue cross is a UrQMD calculation with restricting impact parameter (b) less than 3 fm. Comparing it with the gold line in the most central centrality bin, there might be a residual effect from volume fluctuation. Due to the strong model dependence of the volume fluctuation correction, this correction was not used in the final result shown in Fig.~\ref{fig:energy}.

%Fig.~\ref{fig:rapidity} shows the acceptance dependence of the proton cumulant ratios $C_2/C_1$, $C_3/C_2$, and $C_4/C_2$ for 0-5\% (black squares) and 50-60\% (blue triangles) Au+Au collisions. The respective UrQMD calculations from the two centralities are shown with gold and blue lines. The cumulant ratios in 0-5\% centrality are close to unity (Poisson baseline) in the smallest acceptance window while they seem to saturate in large acceptance. For 50-60\% centrality, cumulant ratios grow with the increase of the acceptance window. Overall, the acceptance dependence trends in 0-5\% and 50-60\% are both well reproduced by the UrQMD calculations.

Figure \ref{fig:energy} shows the collision energy dependence of cumulant ratio $C_4/C_2$ of net-proton and proton multiplicity distributions in central Au+Au collisions~\cite{STAR:2021fge}. %The new measurement of 3 GeV with $-0.5<y<0$ is shown with cyan-filled square. 
As reported in Refs.~\cite{STAR:2021iop,STAR:2020tga} the net-proton and proton $C_4/C_2$ show a non-monotonic energy trend in central Au+Au collisions. A minimum is seen at around $\sqrt{s_{\rm NN}}$ = 20 GeV and then $C_4/C_2$ becomes close to unity with large statistical uncertainty when decreasing collision energy.  
The new measurement of proton $C_4/C_2$ for $\sqrt{s_{\rm NN}}$ = 3 GeV central Au+Au collisions is around -1, which is reproduced by the hadronic transport model UrQMD while at higher energies the non-monotonic energy dependence is not reproduced by various non-critical models including the UrQMD and HRG~\cite{Garg:2013ata} models.
Precision data in the energy window of 3 $<\sqrt{s_{\rm NN}}<$ 20 GeV are needed in order to explore the possibility of critical phenomena.
The HADES collaboration has reported the proton cumulant ratio in $\sqrt{s_{\rm NN}}$ = 2.4 GeV Au+Au collisions within acceptance window $0.4 < p_{T} < 1.6$ GeV/$c$ and $|y|<0.4$: $C_4/C_2$ = 0.15 $\pm$ 0.9 (stat.) $\pm$ 1.4 (sys.)~\cite{HADES:2020wpc} which is consistent with 3 GeV result within uncertainty although detailed comparison should be done within same acceptance.

\section{Summary}
In this paper, we reported cumulant ratios of proton multiplicity distributions in $\sqrt{\rm s_{NN}}$ = 3 GeV Au+Au collisions by the STAR fixed-target experiment. The proton $C_4/C_2$ is observed to be -0.85 $\pm$ 0.09 (stat.) $\pm$ 0.82 (sys.) in the most central 0-5\% centrality at $\sqrt{s_{\rm NN}}$ = 3 GeV. Compared to higher energy results and the transport model calculations, the suppression in $C_4/C_2$ is consistent with fluctuations driven by baryon number conservation and indicates an energy regime dominated by hadronic interactions, which implies that the QCD critical point could exist at energies higher than 3 GeV if discovered in heavy-ion collisions. New data sets have been collected during the second phase of the RHIC beam energy scan program for Au+Au collisions at $\sqrt{s_{\rm NN}}$ = 3 -- 19.6 GeV. The analysis on those datasets will be crucial in exploring the QCD phase structure at high baryon density region and locating the critical point.

\section*{Acknowledgments}
This work was supported by the National Key Research and Development Program of China (Grant No. 2020YFE0202002 and 2018YFE0205201), the National Natural Science Foundation of China (Grant No. 12122505 and 11890711) and the Fundamental Research Funds for the Central Universities(CCNU220N003).

\bibliographystyle{unsrt}

\bibliography{ref_yu_reduced}

\end{document}